# Reversible tuning of magnetic order and intrinsic superconductivity in strained FeTe films via stoichiometry control

*Hao Xu*[1,2,3,†], *Jing Jiang*[4,5,†], *Xuesong Gai*[1,2,3], *Rui-Qi Cao*[1,2,3], *Kaiwei Chen*[1,2,3], *Xiao-Xiao Man*[6], *Haicheng Lin*[1], *Peng Deng*[1], *Ke He*[1,7,8], *Kai Liu*[4,5] *, *Dapeng Zhao*[1] *, *Zhong-Yi Lu*[4,5], *Kai Chang*[1] * & *Chong Liu*[1] *

[1] *Beijing Key Laboratory of Fault-Tolerant Quantum Computing, Beijing Academy of Quantum Information Sciences, Beijing 100193, China*

[2] *Beijing National Laboratory for Condensed Matter Physics, Institute of Physics, Chinese Academy of Sciences, Beijing 100190, China*

[3] *University of Chinese Academy of Sciences, Beijing 100049, China*

[4] *School of Physics and Beijing Key Laboratory of Opto-electronic Functional Materials & Micro-nano Devices, Renmin University of China, Beijing 100872, China*

[5] *Key Laboratory of Quantum State Construction and Manipulation (Ministry of Education), Renmin University of China, Beijing 100872, China*

[6] *College of Physics and Electronic Engineering, Shanxi Normal University, Taiyuan 030006, China*

[7] *State Key Laboratory of Low-Dimensional Quantum Physics, Department of Physics, Tsinghua University, 100084 Beijing, China*

[8] *Frontier Science Center for Quantum Information, 100084 Beijing, China*

[†] These authors contributed equally

*e-mail: kliu@ruc.edu.cn; zhaodp@baqis.ac.cn; changkai@baqis.ac.cn; liuchong@baqis.ac.cn

ABSTRACT

FeTe is a prototypical parent compound of iron-based superconductors. While bulk FeTe is non-superconducting with a long-range bicollinear antiferromagnetic order, superconductivity has been achieved in thin films. However, the approaches usually involve complex oxygen incorporation or interfacial effects, the microscopic mechanisms of which remain elusive. Here, we prepare high-purity, bare FeTe thin films on $SrTiO_3$ using molecular beam epitaxy and investigate their magnetic and superconducting states combining both microscopic and macroscopic characterizations. By reducing the interstitial Fe impurities, we successfully suppress the long-range antiferromagnetic order, enhance the quasiparticle coherence, reduce electron doping and induce superconductivity at above 10 K simultaneously. Moreover, this process is readily reversible by tuning the Fe concentration through well-controlled *in-situ* annealing treatments. Theoretical calculations suggest a spin-fluctuated magnetic ground state under tensile strain. Our findings reveal that strained FeTe thin film is intrinsically superconducting and the precise stoichiometry is a key prerequisite. This work provides insights into the competition between magnetism and superconductivity in iron chalcogenides, and supplies methods for developing stable, high-purity superconducting FeTe films.

Iron-based superconductors (IBSCs) represent a major class of unconventional high-temperature superconductors following the discovery of cuprates, with their single-crystal superconducting transition temperature ($T_c$) reaching as high as ~ 55 K.[1, 2] Compared to cuprates, IBSCs exhibit distinct characteristics such as metallic parent states, multiband electronic structures, and unique pairing mechanisms, which have stimulated sustained research in high-temperature superconductivity.[3, 4]

As a prototypical parent compound of the iron-chalcogenide family, FeTe possesses unique physical properties. It undergoes a structural phase transition near 70 K, accompanied by the formation of a long-range bicollinear antiferromagnetic (BiAFM) order.[5] Unlike most IBSCs, bulk FeTe crystal remains non-superconducting under ambient pressure.[6] However, its ground state is believed to be in close proximity to a superconducting phase, rendering it exceptionally sensitive to external perturbations such as chemical doping[7] and epitaxial strain.[8, 9] This sensitivity makes FeTe an ideal platform for exploring the origins of superconductivity and the complex interplay between magnetism and Cooper pairing. Suppressing the AFM order to induce superconductivity can directly unveil the delicate balance between competing phases, which is crucial for understanding the pairing mechanisms in IBSCs and other unconventional superconductors.

In bulk FeTe, non-stoichiometry manifests as excess Fe ($Fe_{1+y}Te$, where $0.04 < y < 0.17$), where structure, magnetic order and resistivity properties are dependent on $y$.[10, 11] Superconductivity induced directly into FeTe is observed primarily in thin films through two pathways: oxygen incorporation and interfacing with tellurides. FeTe films grown in oxygen or exposed to air/oxygen for extended periods can exhibit superconductivity around 10 K.[12, 13] Similar superconductivity has been observed in the heterostructures between FeTe and a number of various tellurides,[7] such as $Bi_2Te_3$,[14, 15] Cr-doped $(Bi,Sb)_2Te_3$,[16] CdTe.[9] It is hypothesized that interfacial charge transfer,[17] cationic doping,[7] and strain[9] may suppress the AFM order in FeTe, thereby promoting superconductivity.

Despite years of research, the physical mechanisms behind these two phenomena remain unresolved, partly due to experimental obstacles: the oxygen may destroy the surface, and its incorporation with the lattice is complex;[18] the interfacial superconductivity is confined to buried regions, making microscopic characterization difficult. Key questions remain regarding the core factors inducing superconductivity in FeTe and whether a unified mechanism exists for distinct approaches.

In this work, we utilize molecular beam epitaxy (MBE) to precisely control stoichiometry and strain in bare FeTe films on $SrTiO_3$ (STO) substrates. We integrate scanning tunneling microscopy (STM), angle-resolved photoemission spectroscopy (ARPES), *in-situ*/*ex-situ* transport measurements and density functional theory (DFT) calculations to systematically investigate the magnetic and superconducting properties. By reducing the density of interstitial Fe impurities to an extremely low level, we observe that the bicollinear AFM order fades away and superconducting transition at around 10 K emerges in absence of oxidization or heterostructure interface. This process is reversible by tuning the interstitial Fe concentration. The findings indicate that high-purity FeTe thin films could be intrinsically superconducting with the assistance of tensile strain. We identify the stoichiometry and lattice strain as the critical factors for turning AFM state into superconductivity and describe pathways for achieving clean and stable superconducting FeTe films.

## RESULTS AND DISCUSSION

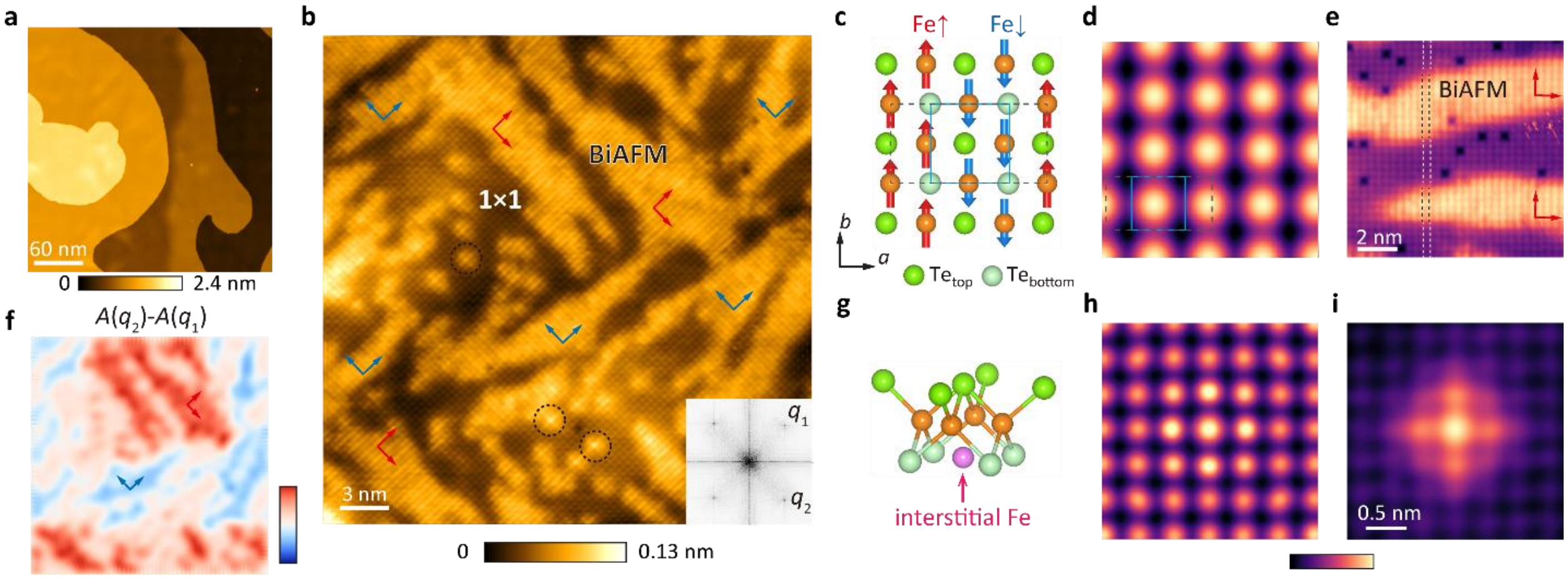

Figure 1. Nanoscale structure and electronic states in FeTe films. (a) STM topographic image of a 20-ML FeTe film (300 nm × 300 nm; setpoint, $V_s$ = 1 V, $I_t$ = 50 pA). (b) Atomically resolved topographic image of the same sample (30 nm × 30 nm; setpoint, $V_s$ = −50 mV, $I_t$ = 500 pA). The red and blue arrows represent the in-plane base vector directions for the BiAFM domains. The dashed circles indicate cross-shaped impurity states attributed to interstitial iron atoms. The inset is the Fourier transform of the image. $q_1$ and $q_2$ are the 1 × 1 Bragg spots. (c) Optimized lattice structure of FeTe in the BiAFM state, obtained by DFT calculations. Red and blue arrows represent the opposite spins on the Fe sites. The magnetic unit cell (dashed box) corresponds to a 2 × 1 supercell of the structural unit cell (solid box). (d) Simulated STM topographic image, which exhibits two-fold symmetry. (e) Nanoscale topographic image showing the shift of the atoms in the BiAFM regions relative to the 1 × 1 regions ($V_s$ = −50 mV, $I_t$ = 500 pA). The dashed lines indicate the arrays of the top Te atoms along the *b* direction. (f) Distribution of the difference of the local amplitude of the two Bragg peaks, extracted by the two-dimensional lock-in technique. (g) Three-dimensional lattice structure with an interstitial Fe atom in between the bottom Te atoms.

(h),(i) Simulated and experimental STM images of the interstitial Fe induced impurity state (experimental setpoint, $V_s = -50$ mV, $I_t = 500$ pA).

We characterize the surface morphology and electronic structure of the FeTe films using STM. Figure 1a shows a large-scale topographic image of a 20-ML FeTe film, which exhibits atomically flat terraces separated by step height of ~0.63 nm, consistent with the FeTe unit cell in the *c* direction.[11] On a finer scale (Figure 1b), the atomically resolved topography reveals a distinct unidirectional stripe-like modulation superimposed on the ordinary 1 × 1 square lattice. These stripes correspond to the BiAFM order, as illustrated in the schematic (Figure 1c), where the bicollinear arrangement of spins breaks the four-fold symmetry. Although the lattice detected by a nonmagnetic tip should maintain the 1 × 1 periodicity,[19] our simulation suggests that the electronic density becomes elongated or more broadened along the *b* direction (Figure 1d and Figure S1), which is confirmed by the zoomed-in experimental image (Figure 1e). More significantly, it also exhibits a relative shift of the Te atom array in the BiAFM region, in agreement with the distortion in our calculated structure in Figure 1c or in the monoclinic phase of the bulk below $T_N$.[9, 10]

The BiAFM order manifests as two types of domains, oriented perpendicular to each other (indicated by red and blue arrows in Figure 1b). To further visualize this local two-fold symmetry, we employed a two-dimensional lock-in technique[20-22] to map the local amplitude difference between the two orthogonal Bragg peaks, $q_1$ and $q_2$. The resulting asymmetry map (Figure 1f) clearly delineates the spatial distribution of these magnetic domains, where red and blue regions represent domains with mutually perpendicular BiAFM wavevectors. This rules out asymmetric tip apex as the origin of the observed stripes. and establishes a direct way to recognize BiAFM order with a nonmagnetic tip. We also acquired STM images with magnetic tips and got the spin-resolved 2 × 1 order in similar BiAFM regions (Figure S2) in line with previous results on bulk FeTe[23].

Furthermore, we observe four-lobed cross-shaped defects (dashed circles in Figure 1b), which center at the top Te sites (Figure 1i). These features are identified as impurity states induced by interstitial iron ($Fe_{int}$) atoms that lie in between the bottom-layer Te atoms and beneath the top layer Te atoms[11](Figure 1g). Our theoretical simulation (Figure 1h and Figure S3a) matches the high-resolution experimental image of a single impurity (Figure 1i and Figure S3c).

Tunneling spectra reveal apparent difference in the electronic density of states (DOS) of the BiAFM and 1 × 1 regions (Figure S4). Furthermore, when measured with larger bias voltage of 500 mV, the BiAFM region and the cross-shaped defects no longer show apparent contrast to the 1 × 1 region (Figure S5), which confirms that these features are mainly contributed by electronic states, and the excess Fe atoms reside underneath the surface layer.

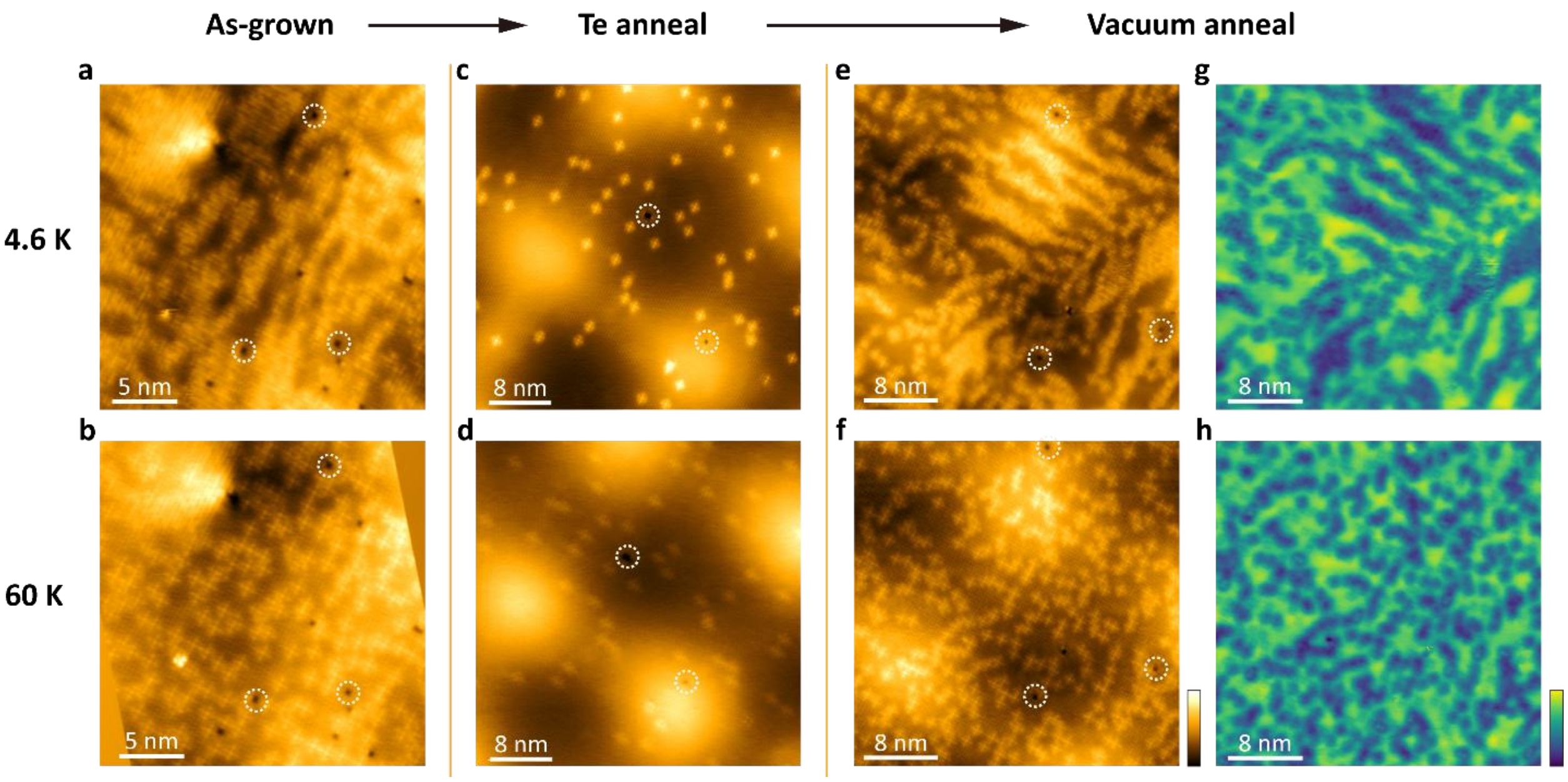

Figure 2. STM topography and electronic evolution of FeTe thin films across annealing cycles. (a, b) Topographic images of the as-grown 10-ML film around the same area, taken at 4.6 K and 60 K, respectively. (c, d) Topographic images of the same sample after Te-vapor anneal. (e, f) Topographic images of the same sample after subsequent vacuum anneal. The dashed circles mark the corresponding point defects in each pair of images, serving as anchor points for location. (g, h) Spatial mapping of d$I$/d$V$ corresponding to (e) and (f), respectively. Affine transformations were applied to all images to correct thermal drift. (Setpoints for all panels: $V_s = -50$ mV, $I_t = 500$ pA)

The ratio between the BiAFM and 1 × 1 regions is flexible. In the 10-ML film grown with lower Te flux, the topography reveals extensive BiAFM regions at 4.6 K (Figure 2a). Upon warming the sample to 60 K close to the Néel temperature, the BiAFM order vanishes, uncovering an amount of $Fe_{int}$ impurities (cross-shaped defects in Figure 2b) that corresponds to a composition of $Fe_{1.028}Te$ in the topmost layer, similar to the common value of the bulk. Following Te-vapor annealing, a dramatic reduction in the density of the $Fe_{int}$ impurities is observed, accompanied by the almost disappearance of the BiAFM regions (Figure 2c,d). The composition is estimated to be $Fe_{1.003}Te$, approaching perfect stoichiometry. This evolution is driven by the chemical potential gradient created by the Te vapor, which extracts and reacts with the $Fe_{int}$ to restore the stoichiometric FeTe phase.

Subsequent vacuum annealing at the growth temperature (Figure 2e,f) evaporates a portion of Te atoms and reintroduces a higher density of $Fe_{int}$, reverting the film to $Fe_{1.029}Te$ in a state dominated by BiAFM order. The d$I$/d$V$ maps (Figure 2g,h) further illustrate the strong spatial modulation of the local DOS. Notably, despite the random distribution of $Fe_{int}$ throughout the layers, its specific electronic signature is screened by the strong DOS of the BiAFM order at low temperatures and only becomes observable when the BiAFM phase vanishes at higher temperatures. These observations reveal that the ground state is highly sensitive to excess Fe and the BiAFM is not the dominant phase in the stoichiometric FeTe films.

| | **NM** | **Néel** | **Stripe** | **Bicollinear** | **Dimer** | **Trimer** |
|---|---|---|---|---|---|---|
| Bulk FeTe | 185.89 | 91.52 | 39.12 | 0.00 | 19.43 | 24.09 |
| Strained bulk FeTe | 177.35 | 64.89 | 15.07 | 19.54 | 0.00 | 4.60 |
| Strained BL FeTe | 182.37 | 60.00 | 12.75 | 13.96 | 0.00 | 0.87 |
| Strained ML FeTe | 192.22 | 59.60 | 14.22 | 16.69 | 0.00 | 2.28 |
| MLFeTe/STO | 177.20 | 51.13 | 20.20 | 17.72 | 0.00 | 2.82 |

Table 1. Comparison of different magnetic configurations in bulk and strained FeTe. Total energies (in units of meV/Fe) from first-principles calculations for NM, Néel AFM, stripe AFM, bicollinear AFM, dimer AFM and trimer AFM states in bulk FeTe, strained bulk FeTe ($a = b = 3.905$ Å), strained bilayer FeTe ($a = b = 3.905$ Å), strained monolayer FeTe ($a = b = 3.905$ Å), and monolayer FeTe on STO substrate ($a = b = 3.905$ Å). The energies of corresponding lowest-energy states in bulk and strained FeTe are set to zero.

To understand the nature of the $1 \times 1$ phase, we performed first-principles calculations to compare total energies of different magnetic configurations (Table 1) for bulk FeTe, strained bulk FeTe (with the in-plane lattice constant fixed at 3.905 Å while allowing fully relaxation along the *c* direction), strained bilayer and monolayer FeTe (with the in-plane lattice constant fixed at 3.905 Å), and monolayer FeTe on STO. The structure models are shown in Figure S6. Five typical magnetic configurations were considered (Figure S7), including the Néel AFM state, stripe AFM state, bicollinear AFM state, dimer AFM state,[24] and trimer AFM state (the calculated band structures are displayed in Figure S8).[25] The energy of nonmagnetic (NM) state was also calculated for comparison. For free-relaxed bulk FeTe, the bicollinear AFM state is identified as the ground state with the lowest total energy, consistent with previous study.[5] However, the energies shift significantly when considering tensile strain, since the exchange coupling is sensitive to the Fe-Fe distance and the Fe-Te-Fe angle. In the cases of all the strained systems, the dimer AFM state emerges as the most stable configuration. Remarkably, the energy difference between the dimer AFM ground state and the trimer AFM state is small, ranging from only 0.87 to 4.60 meV/Fe. This near-degeneracy indicates that these two magnetic configurations are in strong competition. Thermal or quantum fluctuations, as well as local strain variations, can easily destabilize a uniform magnetic phase, preventing the formation of a robust, long-range magnetic order.

This magnetic frustration provides a compelling explanation for experimental observations. In the almost stoichiometric FeTe films with tensile strain from the substrate, the STM topography effectively restores the apparent 1 × 1 structure due to the spin fluctuation. In addition, our experiments demonstrate that the toggling between the BiAFM and magnetically fluctuated state can be readily controlled by the stoichiometry of FeTe, indicating the $Fe_{int}$ plays a role in stabilizing the BiAFM state, in competition with strain. The switching behaviors have no significant thickness dependence for 10–30 ML films in our experiments.

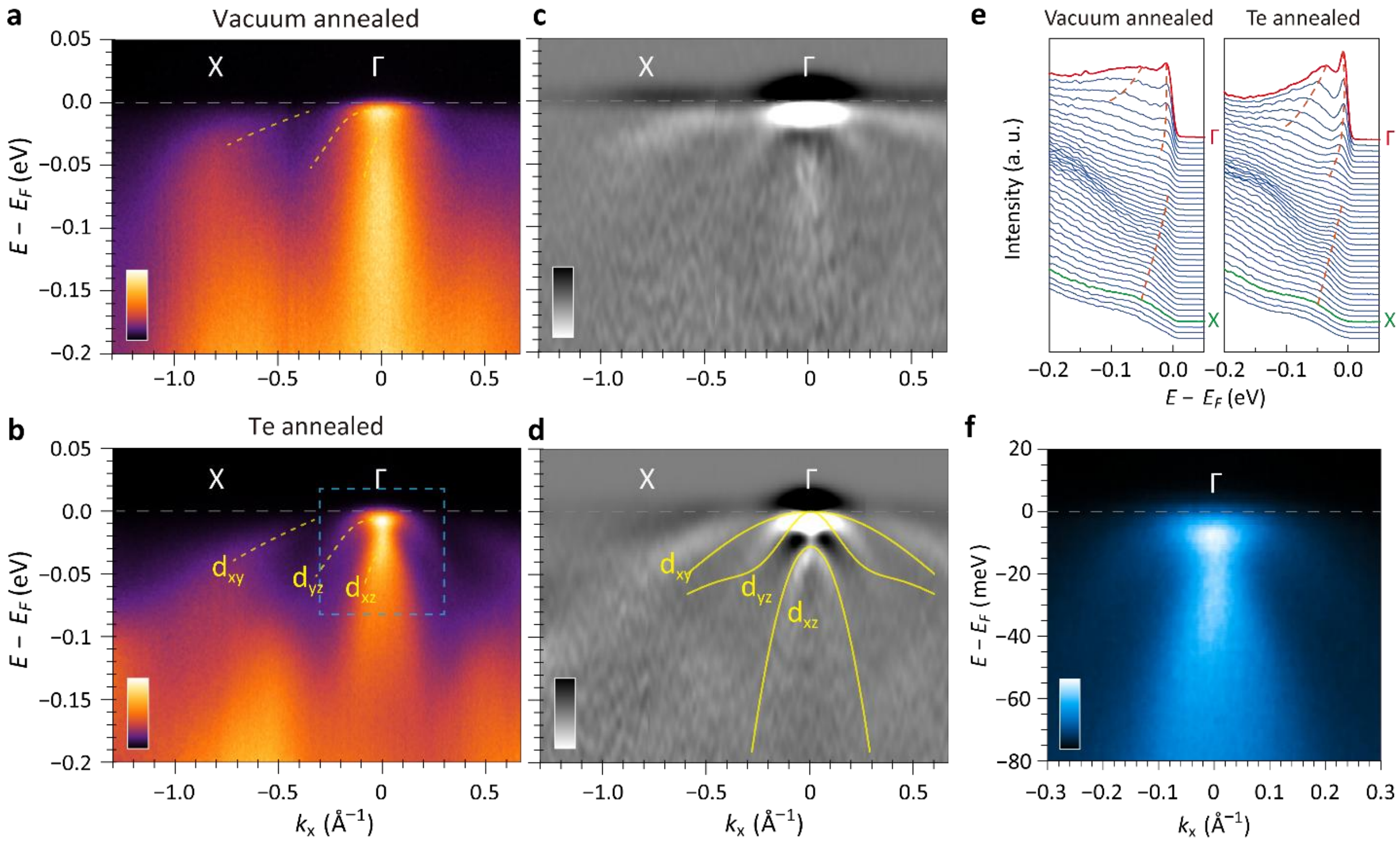

Figure 3. Electronic structure evolution of 10-ML FeTe films measured by ARPES at 5.5 K. (a, b) Energy-momentum spectral maps along the Γ-X direction of the vacuum-annealed and Te-annealed films, respectively. Three hole-type bands are recognized and marked by the dashed lines.[26] (c, d) Second derivative maps corresponding to (a) and (b), respectively. The solid lines in (d) are the DFT band structure of strained bulk FeTe near the Γ point, the $d_{xy}$ and $d_{yz}$ bands are renormalized by different inverse spectral weights ($Z_{xy}^{-1} = 15, Z_{yz}^{-1} = 10$) obtained from DMFT calculations reported in the Ref.[27] (e) Momentum-dependent energy distribution curves (EDCs) from the Γ point to the X point, comparing the vacuum-annealed (left) and Te-annealed (right) states. The dashed lines trace the band dispersion. (f) High-resolution ARPES spectrum near the Fermi level ($E_F$) at the Γ point for the Te-annealed film, corresponding to the range marked by the dashed box in (b).

To further elucidate the impact of excess iron removal on the electronic properties, we performed high-resolution ARPES measurements (Figure 3). In the vacuum-annealed state (Figure 3a), the hole-like bands near the Γ point appear highly incoherent and diffuse. This suppression of quasiparticle coherence is a hallmark of strong scattering induced by the presence of magnetic $Fe_{int}$. The coexistence of the BiAFM domains and the 1 × 1 regions as observed in Figure 2e also leads to strong nonuniformity of electronic states (Figure 2g).

Te-vapor anneal leads to a significant sharpening of the spectral features and an enhancement of the spectral weight near the Fermi level ($E_F$), as shown in Figure 3b,e. The energy distribution curves (EDCs) in Figure 3e clearly demonstrate the emergence of sharp quasiparticle peaks near the Γ point in the Te-annealed state. It was observed in FeTe bulk crystals that Mott localization selectively occurs on the $d_{xy}$ band, leading to extremely strong renormalization and diminished spectral weight $E_F$.[26] Similar decoherence of the $d_{xy}$ band was also reported for vacuum annealed 10-ML FeTe films, consistent with our results in Figure 3a.[27] However, the $d_{xy}$ band of our Te-annealed films show stronger coherence and smaller effective mass (Figure 3d) than the vacuum annealed films (Figure 3c), rather similar to that of the superconducting $FeTe_{1-x}Se_x$,[26, 27] likely due to the removal of Fe impurities and the suppressing of AFM order.

Notably, zoomed-in spectrum of the Te-annealed film (Figure 3f) reveals a possible linear dispersion near the $E_F$ at the Γ point. This Dirac-cone-like structure is further manifested in the second derivative (Figure 3d). It bears a resemblance to the topological surface states observed in the related topological superconductor $FeTe_{1-x}Se_x$.[26, 27] The observation indicates that once the magnetic phase is suppressed, FeTe may also host non-trivial topological electronic states, consistent with the theoretical predictions for iron-chalcogenide superconductors.[28]

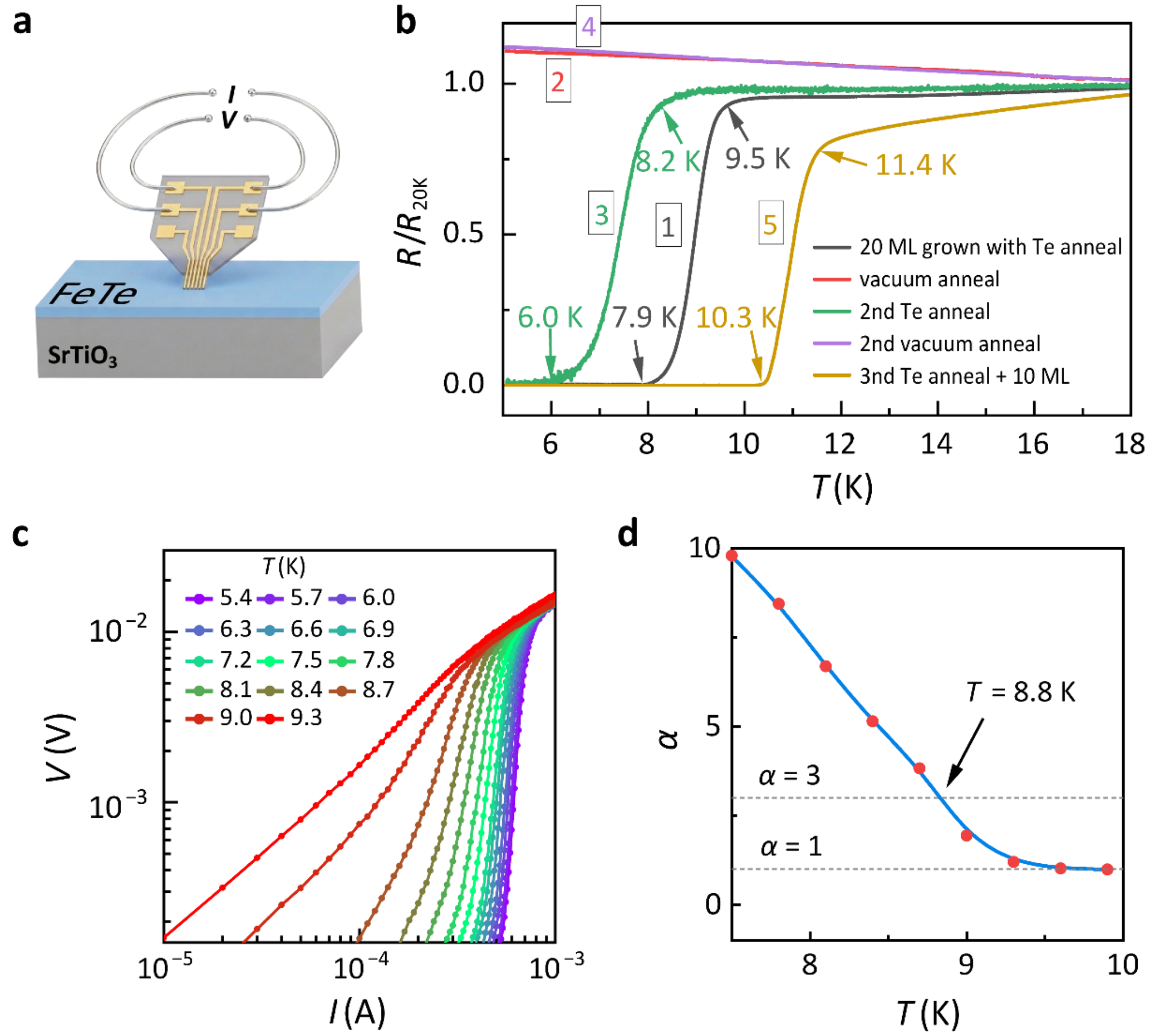

Figure 4. *In-situ* reversible tuning of superconductivity in a FeTe film. (a) Schematic of the microscopic transport measurement setup. (b) Normalized resistance vs. temperature for a FeTe film through sequential annealing cycles. (c) *V-I* characteristics taken on Stage 1 at temperatures from 5.4 K to 9.3 K across the superconducting transition. (d) Temperature dependence of the power-law exponent α deduced from the power-law fits in (c).

To investigate the role of stoichiometry in the emergence of superconductivity, we performed *in-situ* microscopic four-probe transport measurements[29, 30] on FeTe thin films (Figure 4a) subjected to sequential annealing treatments. The sample was first prepared in an ultra-high vacuum (UHV) interconnected MBE chamber through alternating growth (for 30 minutes) and Te-vapor annealing (for 10 minutes) until the nominal thickness reached 20 ML (Stage 1). This recipe is to ensure sufficient reaction of the entire film with Te. Then the sample was immediately measured without air exposure so that the oxygen incorporation is excluded. As shown in Figure 4b, the *R*-*T* curve exhibits a sharp superconducting transition with $T_{c,onset} \approx 9.5$ K and $T_{c,0} \approx 7.9$ K. This is attributed to the effective removal of excess Fe atoms during the Te-vapor treatment, corresponding to the 1 × 1 dominated phase observed in STM (Figure 2c).

The nature of the superconducting transition is further characterized by analyzing the voltage-current (*V*-*I*) curves (Figure 4c). The *V*-*I* curves exhibit nonlinear power-law behavior, $V \propto I^{\alpha}$, characteristic of two-dimensional superconductors. By fitting and plotting the power-law exponents $\alpha$ derived from Figure 4c, we observed a Berezinskii-Kosterlitz-Thouless (BKT) transition[31] (Figure 4d). The transition temperature $T_{BKT}$, defined by the criterion $\alpha = 3$, was determined to be 8.8 K for the 1st Te-annealed state.

To confirm the inhibitory role of Fe interstitials, the film was subjected to vacuum annealing at the growth temperature for 1.5 hours (Stage 2). This process regenerates excess Fe in the lattice, and consequently, superconductivity was entirely suppressed, yielding a metallic resistive profile (Figure 4b). The reversible nature of this phase transition was demonstrated by the subsequent Te-vapor annealing for 40 minutes (Stage 3), which restored the superconducting state ($T_{c,onset} \approx 8.2$ K). Then a vacuum annealing for 90 minutes damaged the superconductivity again (Stage 4). Further enhancement of $T_{c,onset}$ to 11.4 K was observed in Stage 5 after additional Te annealing and further growth of 10 ML, suggesting a thickness-dependent stabilization of the superconducting phase. The evolution of the *in-situ* transport properties through successive annealing cycles provides compelling evidence that the emergence of superconductivity in FeTe is directly governed by the concentration of excess Fe.

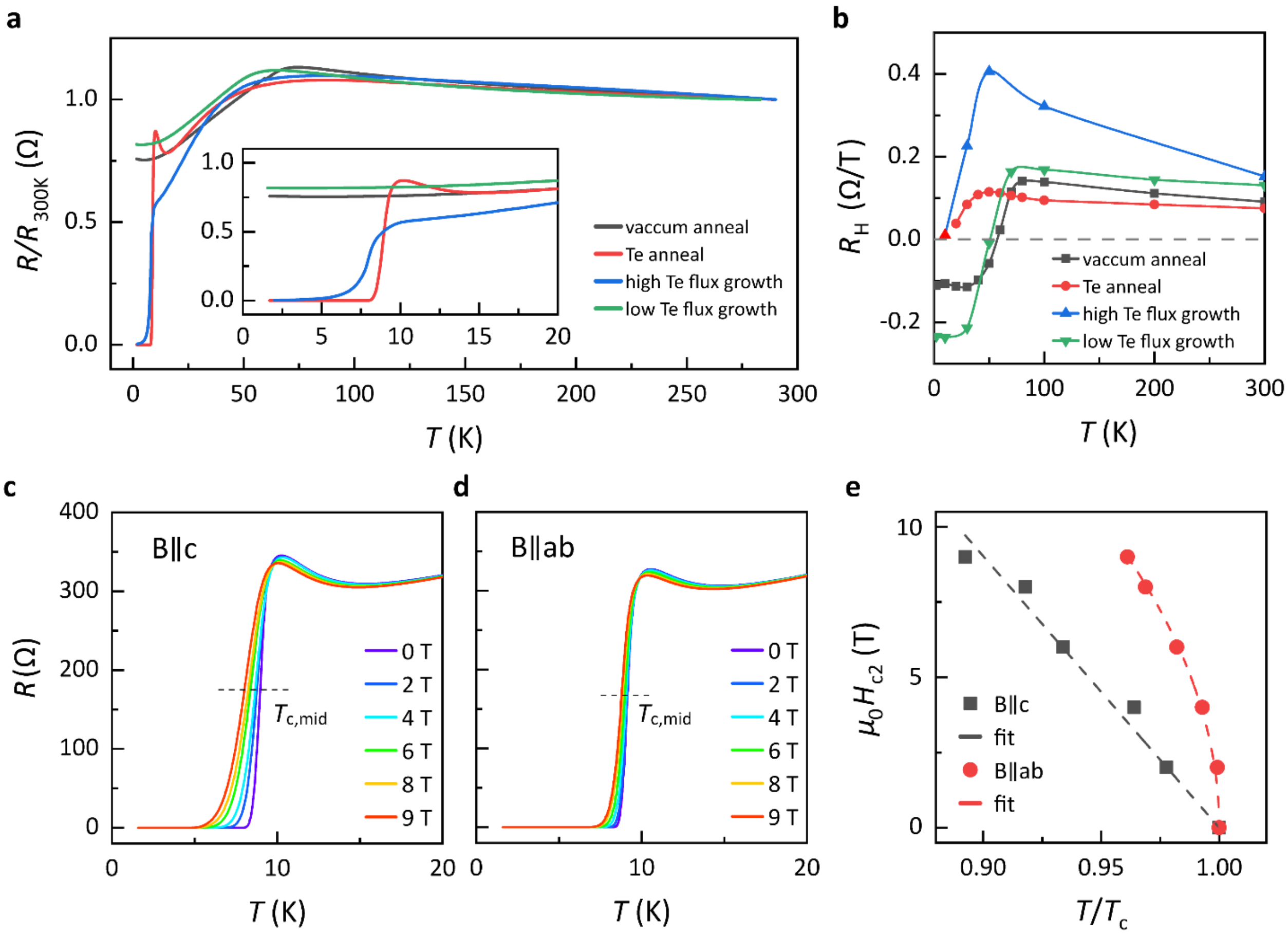

Figure 5. *Ex-situ* characterization of FeTe films prepared with different methods. (a) Normalized resistance vs. temperature for 4 different FeTe samples. Black and red lines: 30-ML films by alternating deposition-annealing in vacuum and Te vapor, respectively. Blue and green lines: 20-ML films by continuous deposition with high and low Te flux, respectively. (b) Temperature-dependent Hall coefficients for the samples in (a). (c, d) *R*-*T* curves around the superconducting transition of the Te-annealed sample, taken with various magnetic fields. (e) Upper critical fields at different temperatures extracted from (c) and (d). The dashed lines are fittings using the GL equations.

To investigate the macroscopic electronic properties and superconductivity, we performed *ex-situ* transport measurements on FeTe films prepared by different growth and annealing protocols (samples were capped by $AlO_x$ to prevent oxidization). As shown in Figure 5a, superconductivity is successfully achieved in films prepared by both alternating deposition-annealing (red line) and continuous deposition (blue line) methods. Again, the emergence of superconductivity is highly sensitive to the tellurium environment during preparation. While the Te-annealed and high-Te-flux growth samples exhibit clear superconducting transitions with $T_{c,onset} \approx 10$ K, samples prepared under vacuum annealing (black line) or low Te flux (green line) remain non-superconducting. The downturns at 60–70 K, which correspond to the known AFM and structure transition[6], are softened for the superconducting samples. This suggests that the AFM state is competing with superconductivity.

The impact of $Fe_{int}$ removal is further evidenced by the temperature-dependent Hall resistance (Figure S9). As shown in Figure 5b, for the superconducting samples (red and blue), the Hall coefficients ($R_H$) remain positive across the entire temperature range until the superconducting transition. This behavior is distinct from the low-Te-flux or vacuum-annealed samples, which typically exhibit a sign reversal in $R_H$ associated with the onset of the AFM state.[32] This suggests that the elimination of $Fe_{int}$ reduces the extrinsic electron-doping effect.

The nature of the superconducting state was further probed under various magnetic fields for the Te-annealed sample (Figure 5c,d). Here, $T_c$ is defined as the temperature at the 50% drop of normal state resistance. There is anisotropy in the upper critical field ($H_{c2}$) where the in-plane field suppresses superconductivity less effectively than the out-of-plane field. We extracted the $H_{c2}$ values and fitted them using the Ginzburg-Landau (GL) equations (Figure 5e),

$$\mu_0 H_{c2}^{\perp}(T) = \frac{\Phi_0}{2\pi\xi_{\mathrm{GL}}^2}\left(1 - \frac{T}{T_c}\right) \qquad \text{(Eq. 1)}$$

$$\mu_0 H_{c2}^{\parallel}(T) = \frac{\Phi_0\sqrt{12}}{2\pi\xi_{\mathrm{GL}} d_{\mathrm{SC}}}\sqrt{1-\frac{T}{T_{\mathrm{c}}}} \quad \text{(Eq. 2)}$$

where $\Phi_0$ is the flux quantum, $\xi_{GL}$ is the GL in-plane coherence length at 0 K, and $d_{sc}$ is the superconducting thickness.[33] The fitting yields $\xi_{GL} \approx 1.9$ nm and $d_{sc} \approx 13$ nm. Notably, the anisotropy is much weaker than that in the atomically thin superconductors such as monolayer FeSe/STO,[34, 35] $WTe_2$[36] and ZrNCl.[33] In contrast to monolayer FeSe/STO, the narrower transition width within 2 K, the weaker anisotropy and the larger superconducting thickness of FeTe films imply that the superconductivity exists in a large part of the films instead of being confined at the surface or interface.

The pursuit of superconductivity in iron chalcogenides has long been hindered by the presence of excess Fe, a common stoichiometric challenge in single-crystal growth.[37] Our results suggest that the key to unlocking superconductivity in this system lies in the successful removal of the interstitial Fe, which serves as a primary prerequisite for suppressing the long-range antiferromagnetic phase and facilitating a superconducting ground state. The impact of $Fe_{int}$ on superconductivity is multi-faceted:

(1) Stabilization of long-range magnetic order. As observed in our STM data (Figure 2), higher concentrations of $Fe_{int}$ are closely associated with robust BiAFM domains. The magnetic moments of these $Fe_{int}$ atoms can interplay with the ordinary Fe in the lattice,[38] likely acting as "magnetic anchors" that stabilize long-range magnetic order. Since superconductivity in iron-based materials typically emerges upon the suppression of static magnetism, the persistence of BiAFM order stabilized by Fe acts as a direct competitor to the superconducting phase.

(2) Magnetic impurity scattering. $Fe_{int}$ acts as a magnetic impurity within the FeTe matrix. These impurities induce strong local scattering of quasiparticles, which is known to be pair-breaking, directly disrupting the formation of Cooper pairs.

(3) Electronic doping. Beyond its magnetic effects, $Fe_{int}$ introduces additional electrons into the system, as predicted by calculations[38] and confirmed by Hall effect (Figure 5b). If the superconductivity in FeTe is related with hole doping as implied in previous studies,[17, 39] this unintentional electron doping could shift the chemical potential away from the carrier concentrations required for the emergence of superconductivity.

In addition to the chemical control of $Fe_{int}$, our analysis based on DFT calculations (Table 1) highlights the critical role of epitaxial strain provided by the STO substrate. The energetic competition destabilizes the long-range AFM phase, which resembles the situation in bulk FeSe, where the AFM order is also absent but the spin correlation is strong.[25] By suppressing static magnetic order and enhancing spin fluctuations, the strain creates an environment favorable for Cooper pairing in FeTe, making it more analogous to other iron chalcogenides (even having similar $T_c$). The suppression of structural distortion by strain may also contribute.[9] This could also explain the PLD-grown superconducting films where the lattice shrinks out of plane[8]. Consequently, the synergy between eliminating $Fe_{int}$ and leveraging substrate strain represents the most viable pathway for achieving high-temperature superconductivity in FeTe thin films.

Common mechanisms might exist in the cases of oxygen treatment and telluride interface. It has been suggested in the films grown by pulse laser deposition that the oxygen annealing can reduce the excess Fe[40] and suppress the AFM order.[18] For the FeTe-telluride heterostructures, the usually high Te flux during the subsequent growth of the top layers could have a similar effect to Te-vapor anneal.[41] Therefore, the contribution of interfacial effects in those systems needs to be reexamined. Different experimental approaches to the FeTe-related superconductivity are likely to have universal control factors in essence.

After the submission of this manuscript, a related study was reported by Yan *et al.,*[42] which also identifies stoichiometry as a governing factor for superconductivity in FeTe. Our findings are highly consistent with their observations. However, our work provides additional insights by demonstrating the *in-situ* reversible modulation between the BiAFM and the superconducting state, and revealing their differences in band structure and carrier types. Moreover, we propose that the strain is another tuning factor for the antiferromagnetism and superconductivity, suggesting a complex interplay between strain and stoichiometry that invites further exploration.

## CONCLUSIONS

In summary, this research reveals the interplay between stoichiometry, strain, and magnetism that governs the electronic ground state of FeTe films. By combining atomic-scale imaging with ARPES, transport measurements and theoretical analysis, we demonstrate that the transition from a long-range AFM metal to a superconductor is fundamentally driven by the elimination of interstitial Fe and the exploitation of substrate-induced strain. This work describes a method for synthesizing high-quality superconducting FeTe films and indicates the delicate balance between competing phases in unconventional superconductors.

## METHODS

**Sample preparation.** FeTe films were prepared in an MBE system. The base pressure of the chamber was ~$1 \times 10^{-10}$ Torr. For STM, ARPES and *in-situ* transport measurements, 0.5 wt.%

Nb:STO substrates were annealed successively at 1050 °C for 40 minutes and 1150 °C for 20 minutes to obtain atomically flat surfaces. For *ex-situ* transport measurements, undoped STO substrates were treated successively by water-boiling for 1 hour, HCl etching for 45 minutes and annealing in oxygen at 1120°C for 3 hrs. FeTe films were grown by coevaporating high-purity Fe and Te from standard Knudsen cells while the substrate was held at 270–280 °C. The growth rate was ~0.22 layer/minute. The stoichiometry was adjusted by either tuning the flux of Te or post-annealing in vacuum/Te vapor at the growth temperature. Reflection high-energy electron diffraction (RHEED) was used to monitor the growth progress and sample quality with the electron energy of 10 keV. The diffraction patterns had no obvious change during annealing (see Figure S10). For *ex-situ* transport measurements, the samples were further capped with 2-nm Al films at room temperature, which naturally form $AlO_x$ capping layers in the air.

**STM measurements.** *In situ* STM measurements were conducted on a SI-STM-4K instrument (CIENSS Co., Ltd.). The sample temperature was 4.6 K with liquid helium cooling unless otherwise noted. A polycrystalline PtIr tip was used and calibrated on Ag islands before STM experiments. Images are scanned in the constant current mode. d$I$/d$V$ maps were acquired by scanning the image while adding oscillation on the bias voltage and extracting the demodulation from the tunneling current by a standard lock-in technique. The typical oscillation frequency is 973.2 Hz and the amplitude is 5 mV.

**ARPES measurements.** ARPES measurements were carried out at BAQIS on a lab-based system. The samples were grown in a UHV interconnected MBE system. The ARPES chamber is equipped with a helium discharge lamp (Fermi Instruments) and a DA30-L electron energy analyzer (Scienta Omicron). The photon energy is 21.2 eV. The combined energy and angle resolutions are better than 9.5 meV and 0.5 degrees. The base temperature of the sample was 5.5 K. The pressure of the chamber was below $2 \times 10^{-10}$ Torr during measurements.

***In-situ* transport measurements.** *In-situ* electrical transport measurements were conducted in a home-built UHV system equipped with a piezo-driven micro-four-point probe (M4PP) setup.[29, 30] The width of each probe is 7 μm while the distance between adjacent probes is 3 μm. The sample stage was cooled with liquid helium. The samples were grown in a UHV interconnected MBE system. The data were acquired under a built-in Pulse Delta measurement mode of Keithley Source Meters 6221/2182A with a pulsed current $I = 2$ μA applied.

***Ex-situ* transport measurements.** Magneto-electrical transport properties were measured in the four-terminal method in a physical property measurement system (Quantum Design) and a CPMS system (CSIC PRIDe Cryogenic Technology). Al wires were cold bonded on the samples with indium lumps as the electrodes.

**First-principles calculations.** The magnetic properties and STM simulations of FeTe bulk and films were studied by using the density functional theory calculations as implemented in the VASP package.[43-45] The fully spin-polarized electronic structure calculations were carried out by using the projector augmented wave (PAW) method.[46] The generalized gradient approximation (GGA) of Perdew-Burke-Ernzerhof (PBE) type[47] for the exchange-correlation potentials was adopted. The kinetic energy cut-off of plane wave basis was set to 520 eV. The convergence criterion for the forces on all atoms was set to 0.01 eV/Å. The DFT-D3 method[48] was used to describe the van der Waals (vdW) interaction between the adjacent FeTe layers except for the monolayer FeTe. The STM simulations were performed to investigate the surface electronic structure of FeTe in both the bicollinear AFM and nonmagnetic states. The simulated STM images were generated by visualizing the charge density integrated within a specified energy window (bias voltage referring to Fermi level) using VESTA.[49] For the case with interstitial Fe located beneath the surface Te atom, the Fermi level used in the STM simulation was shifted to $E_F$ + 0.13 eV, where $E_F$ is the Fermi level obtained from self-consistent calculation. This adjustment was introduced to account for the substrate- or $Fe_{int}$-induced electron doping effect and the correlation effect not included in DFT calculation.[39, 50]

ASSOCIATED CONTENT

**Supporting Information**

The following file is available.

Structure models and spin configurations for calculations, calculated band structures, additional STM/STS and transport data, RHEED images. (PDF)

AUTHOR INFORMATION

**Corresponding Author**

Kai Liu – Email: kliu@ruc.edu.cn

Dapeng Zhao – Email: zhaodp@baqis.ac.cn

Kai Chang – Email: changkai@baqis.ac.cn

Chong Liu – Email: liuchong@baqis.ac.cn

**Author Contributions**

C.L. conceived and designed the research. H.X. and X.G. prepared the samples. H.X., X.G., R.-Q.C. and C.L. conducted the STM measurements. H.X. and X.G. conducted ARPES measurements.

H.X., D.Z. and K.Chen conducted transport measurements. J.J., X.-X.M., K.L. and Z.-Y.L. carried out first principles calculations. H.X. and C.L. analyzed the data. H.X., D.Z., X.G., R.-Q.C., H.L. and C.L. maintained the equipment. K.Chang and C.L. supervised the project. H.L., P.D., K.H., K.L., Z.-Y.L. and K.Chang discussed the interpretation of results. H.L., P.D., K.L., Z.-Y.L., D.Z., K.Chang and C.L. offered funding support. H.X. and C.L. wrote the manuscript with the input from all authors.

**Notes**

The authors declare no competing financial interest.

ACKNOWLEDGMENT

This work was supported by Quantum Science and Technology-National Science and Technology Major Project (2023ZD0300500), National Natural Science Foundation of China (Grant Nos. 12204046, 12304189, 12204048, 12304208), Beijing Natural Science Foundation (Grant Nos. 1252036, 1242037, 1232035), Beijing Municipal Science & Technology Commission (Grant No. Z221100002722013), Open Research Fund Program of the State Key Laboratory of Low-Dimensional Quantum Physics (Grant No. KF202508). The theoretical work was supported by the National Key R&D Program of China (Grants No. 2022YFA1403103 and No. 2024YFA1408601) and the National Natural Science Foundation of China (Grants No. 12174443 and No. 12434009). Computational resources have been provided by the Physical Laboratory of High Performance Computing at Renmin University of China and the Beijing Super Cloud Computing Center.

Supporting Information

# Reversible tuning of magnetic order and intrinsic superconductivity in strained FeTe films via stoichiometry control

Hao Xu[1,2,3,†], Jing Jiang[4,5,†], Xuesong Gai[1,2,3], Rui-Qi Cao[1,2,3], Kaiwei Chen[1,2,3], Xiao-Xiao Man[6], Haicheng Lin[1], Peng Deng[1], Ke He[1,7,8], Kai Liu[4,5]*, Dapeng Zhao[1]*, Zhong-Yi Lu[4,5], Kai Chang[1]* & Chong Liu[1]*

[1] *Beijing Key Laboratory of Fault-Tolerant Quantum Computing, Beijing Academy of Quantum Information Sciences, Beijing 100193, China*

[2] *Beijing National Laboratory for Condensed Matter Physics, Institute of Physics, Chinese Academy of Sciences, Beijing 100190, China*

[3] *University of Chinese Academy of Sciences, Beijing 100049, China*

[4] *School of Physics and Beijing Key Laboratory of Opto-electronic Functional Materials & Micro-nano Devices, Renmin University of China, Beijing 100872, China*

[5] *Key Laboratory of Quantum State Construction and Manipulation (Ministry of Education), Renmin University of China, Beijing 100872, China*

[6] *College of Physics and Electronic Engineering, Shanxi Normal University, Taiyuan 030006, China*

[7] *State Key Laboratory of Low-Dimensional Quantum Physics, Department of Physics, Tsinghua University, 100084 Beijing, China*

[8] *Frontier Science Center for Quantum Information, 100084 Beijing, China*

[†]These authors contributed equally

*e-mail: kliu@ruc.edu.cn; zhaodp@baqis.ac.cn; changkai@baqis.ac.cn; liuchong@baqis.ac.cn

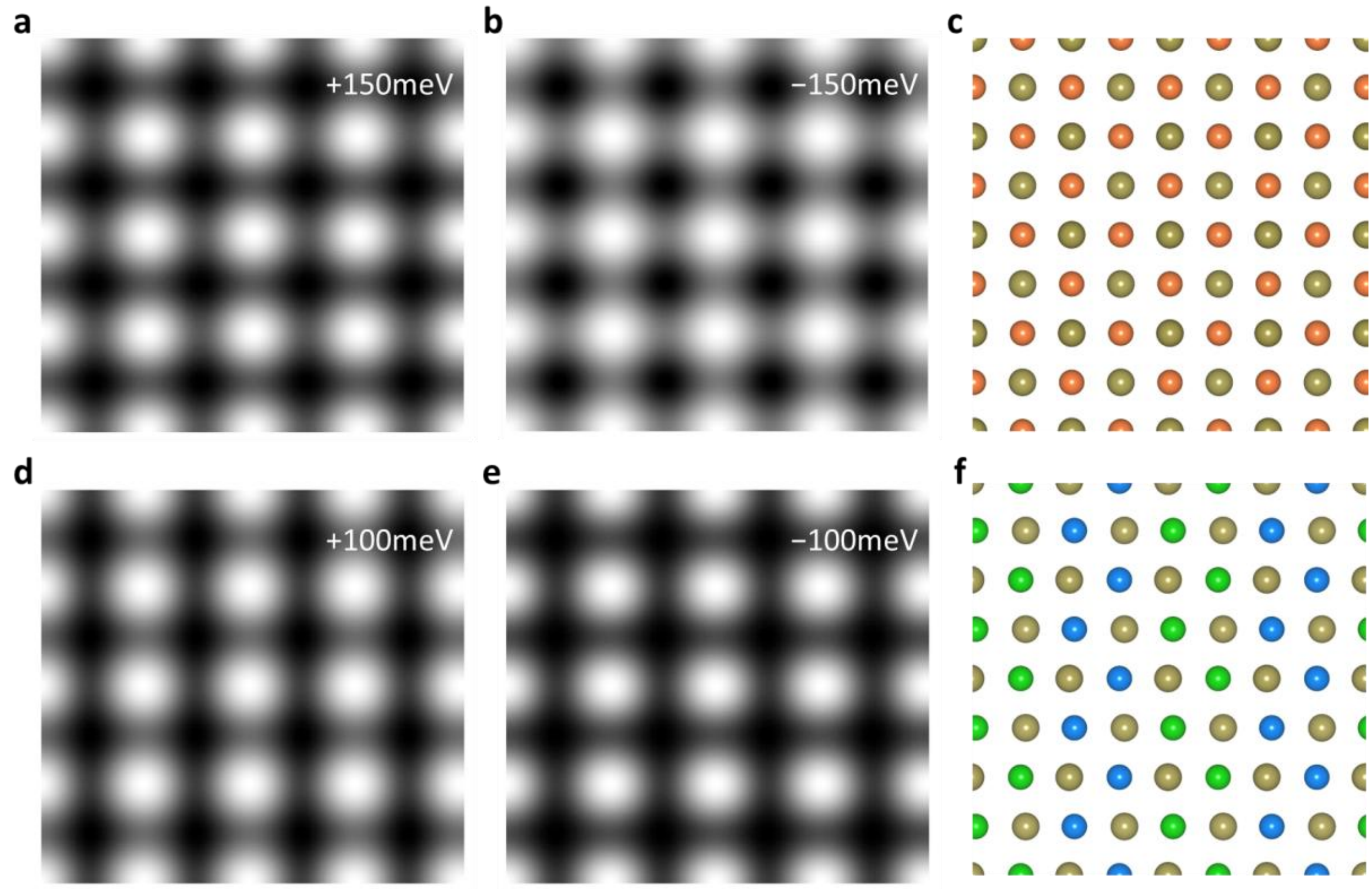

Figure S1. Scanning tunneling microscope (STM) simulations of strained monolayer FeTe. (a, b) Simulated STM image under 150 mV and −150 mV bias in nonmagnetic state, showing a 1×1 periodicity. (c) Top view of the corresponding atomic structure, where brown and orange spheres represent Te and Fe atoms, respectively. (d, e) Simulated STM image under 100 mV and −100 mV bias in bicollinear AFM state. (f) Top view of the corresponding atomic structure, where blue and green spheres denote Fe atoms with opposite spin orientations.

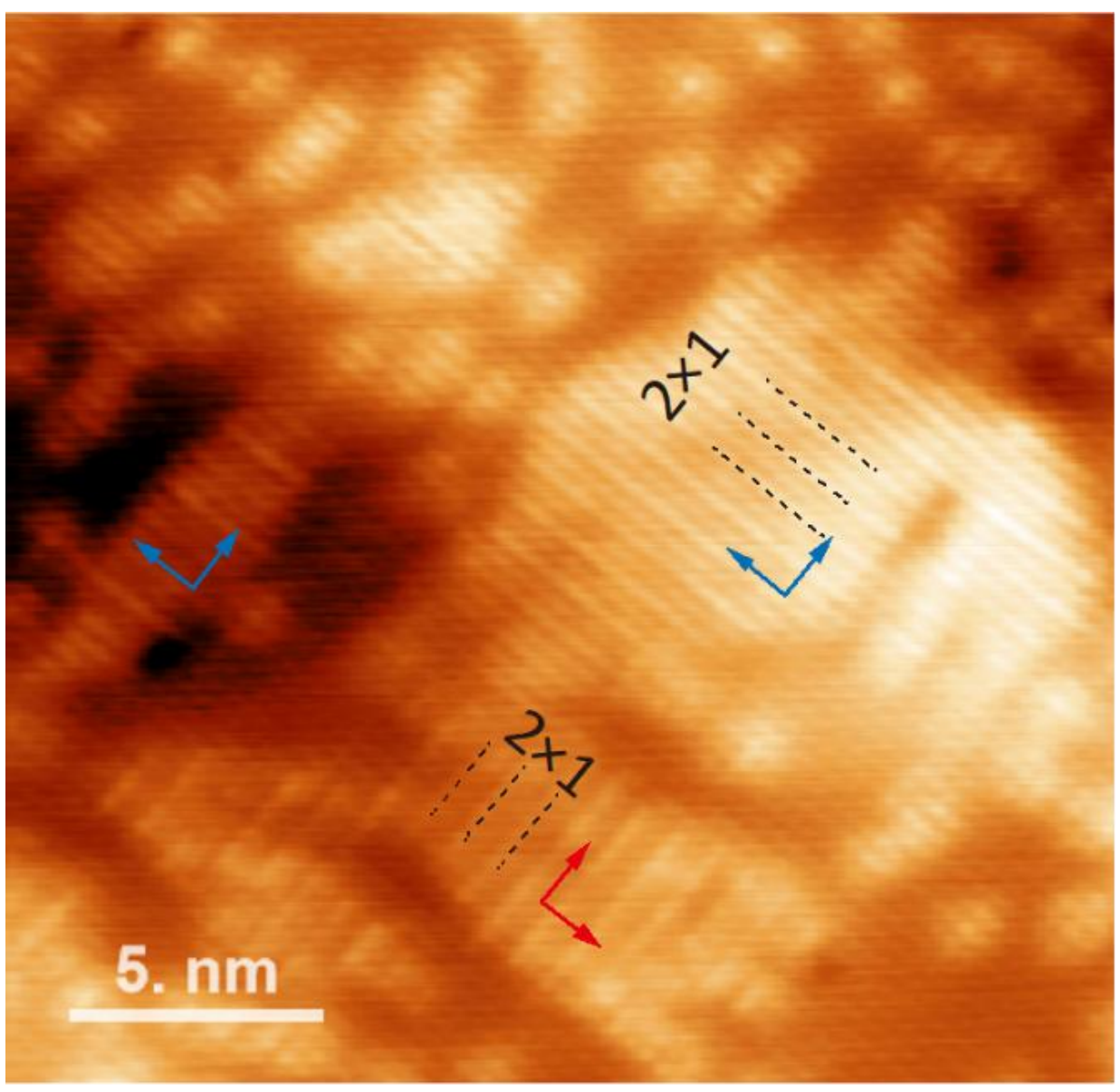

Figure S2. STM image taken with a spin-polarized tip apex ($V_s = -100$ mV, $I_t = 500$ pA), showing domains of 2 × 1 ordering.

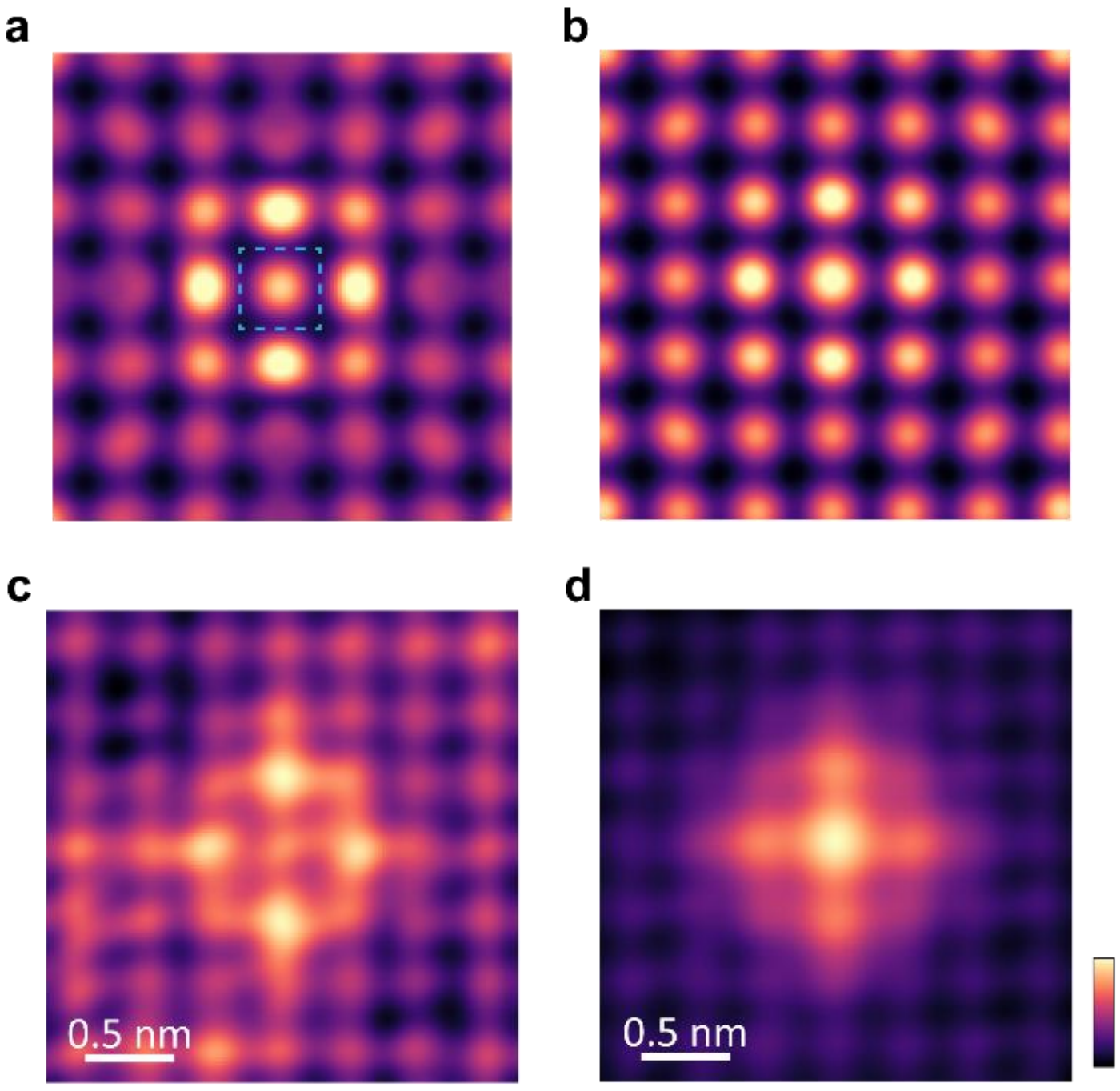

Figure S3. Simulated and experimental STM images of the nonmagnetic FeTe surface with the interstitial Fe. (a) Simulated STM image under positive bias (+220 mV). The Te atom locating above the interstitial Fe is highlighted by the dashed box. (b) Simulated STM image under negative bias (−100 mV). (c, d) Experimental STM images taken under positive ($V_s$ = 50 mV, $I_t$ = 500 pA) and negative bias ($V_s$ = −50 mV, $I_t$ = 500 pA). (b) and (d) are the same as Figure 1h,i in the main text.

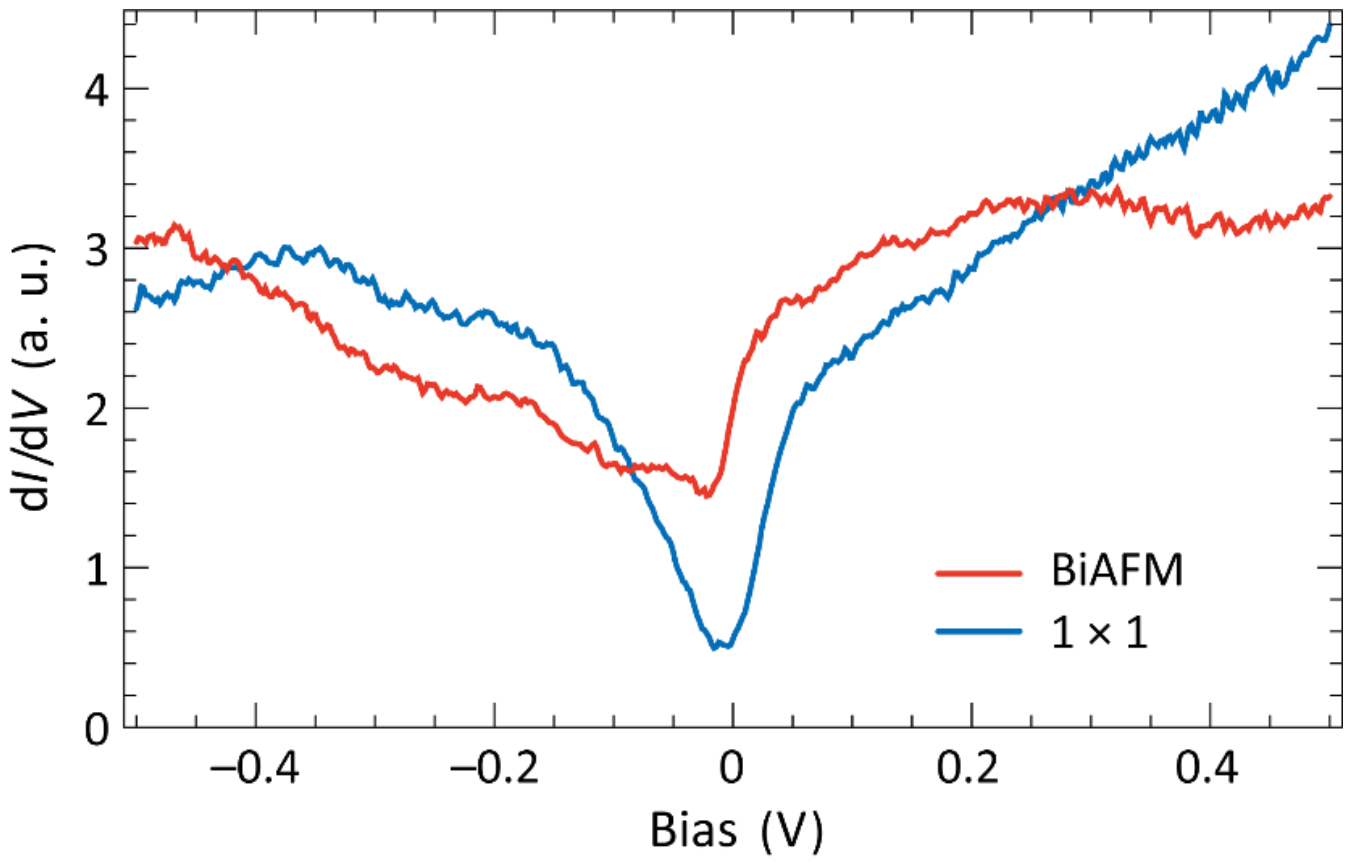

Figure S4. Typical tunneling spectra taken on the BiAFM and 1×1 surface regions ($V_s$ = 500 mV, $I_t$ = 500 pA).

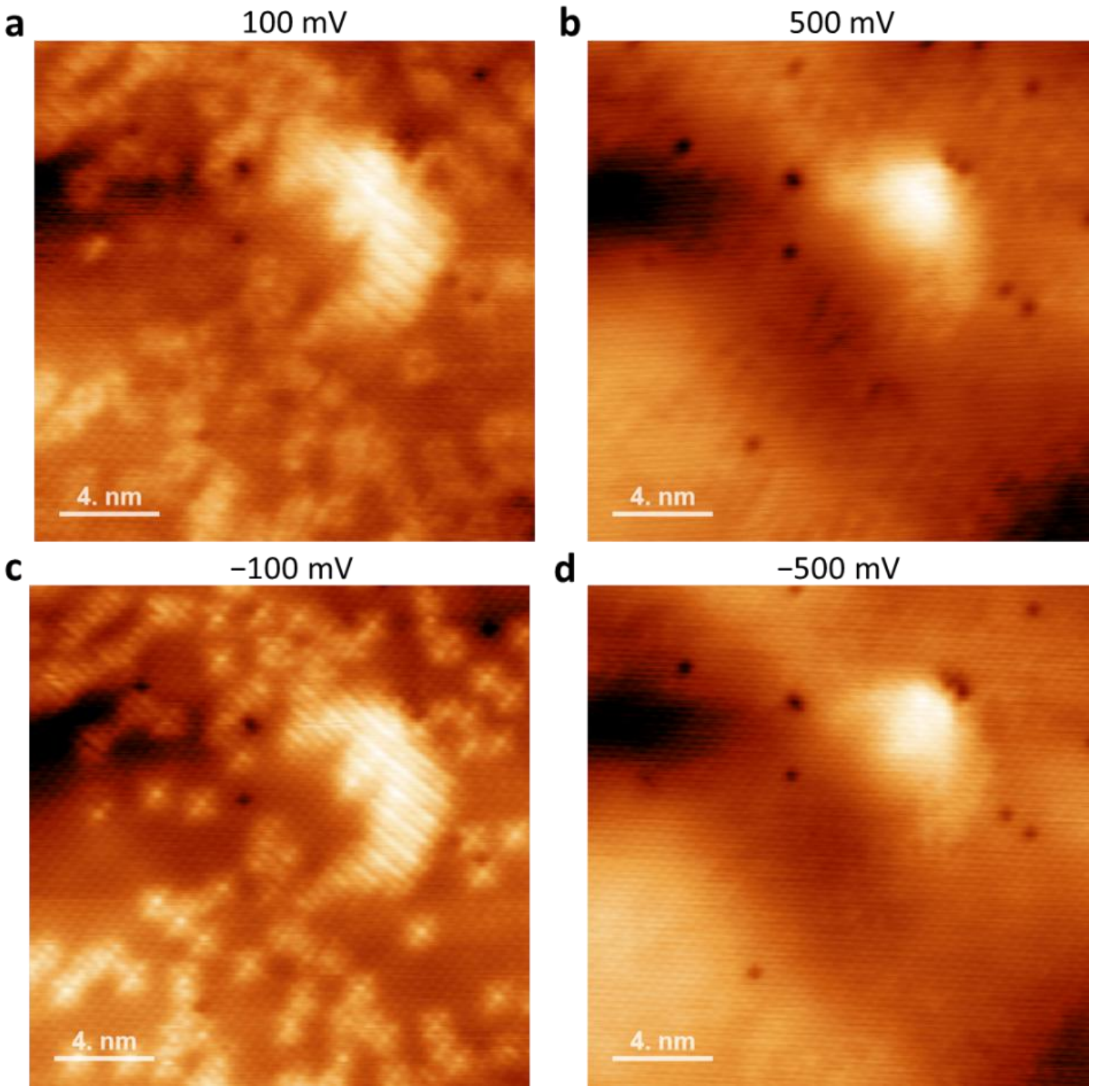

Figure S5. STM image taken under various sample bias voltages ($I_t$ = 500 pA). Both the BiAFM stripe order and the cross-shaped impurity states vanish at higher bias.

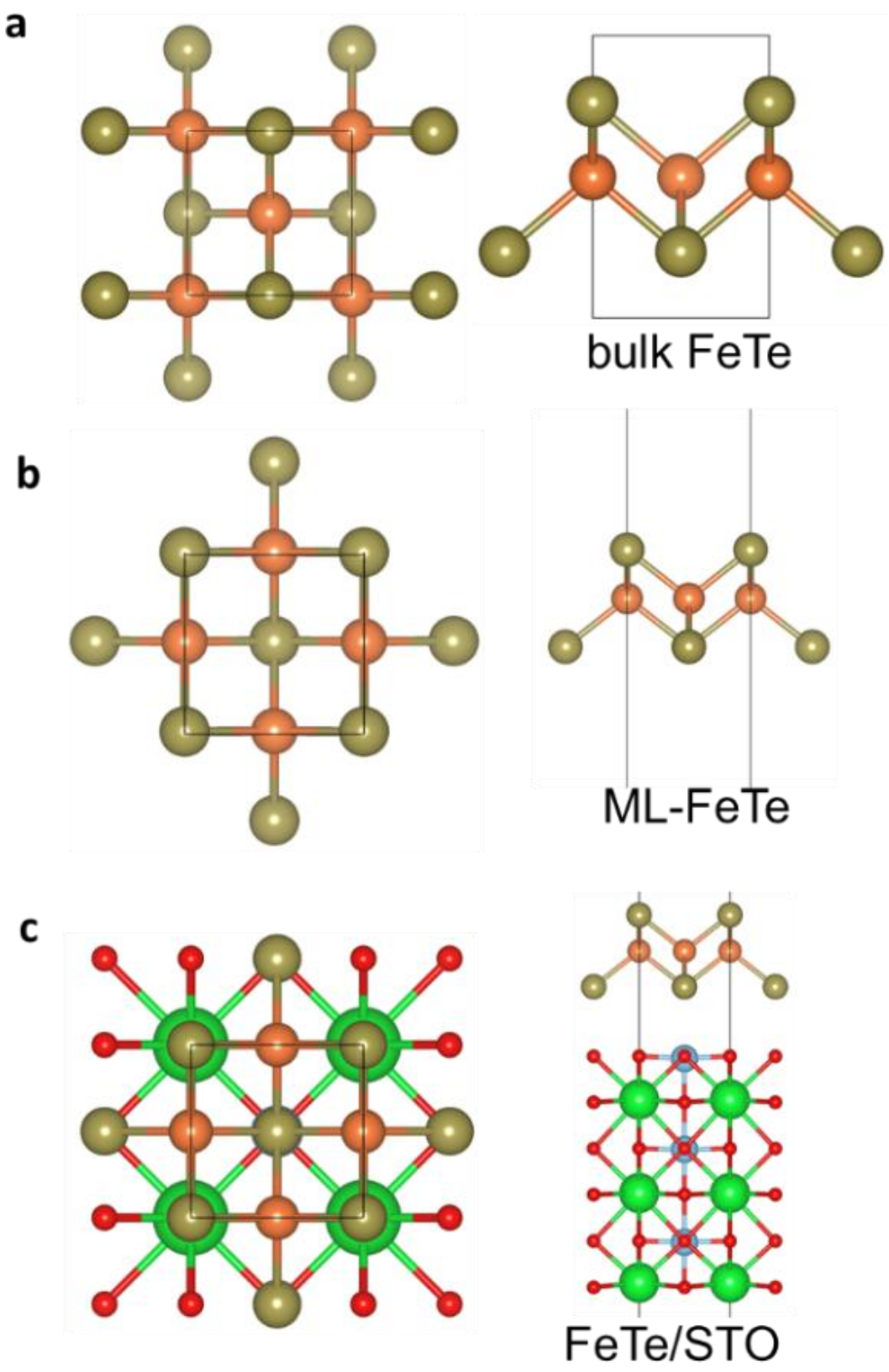

Figure S6. Top and side views of different FeTe structures. (a) Bulk FeTe, (b) strained monolayer FeTe, and (c) monolayer FeTe on $SrTiO_3$ substrate (FeTe/STO).

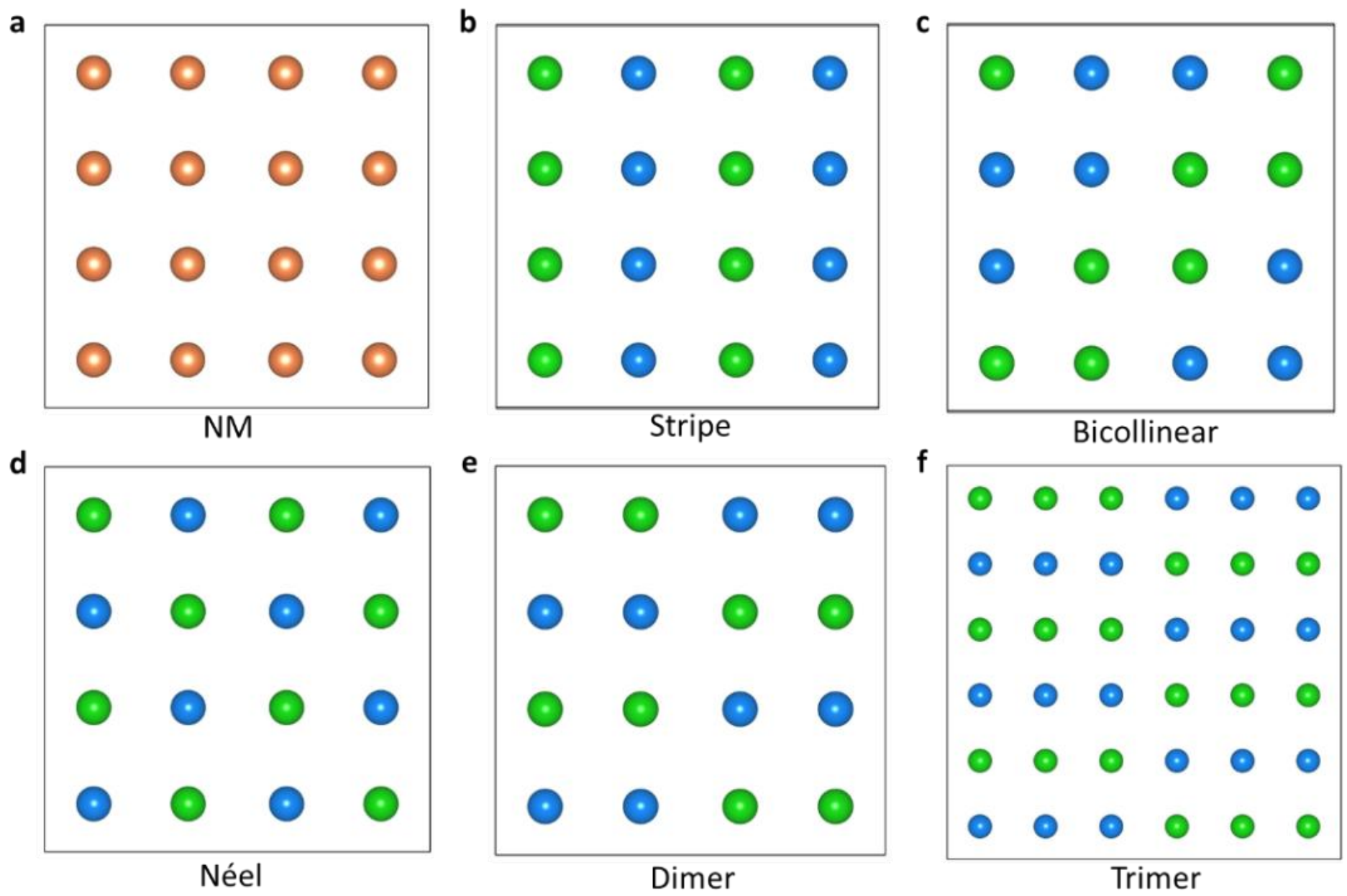

Figure S7. Typical magnetic configurations of the Fe-Fe square lattice in FeTe. (a) Nonmagnetic state (NM). (b) Stripe antiferromagnetic (AFM) state (Stripe). (c) Bicollinear AFM state (Bicollinear). (d) Néel AFM state (Néel). (e) Dimer AFM state (Dimer). (f) Trimer AFM state (Trimer). The green and blue spheres represent Fe atoms with opposite spin orientations.

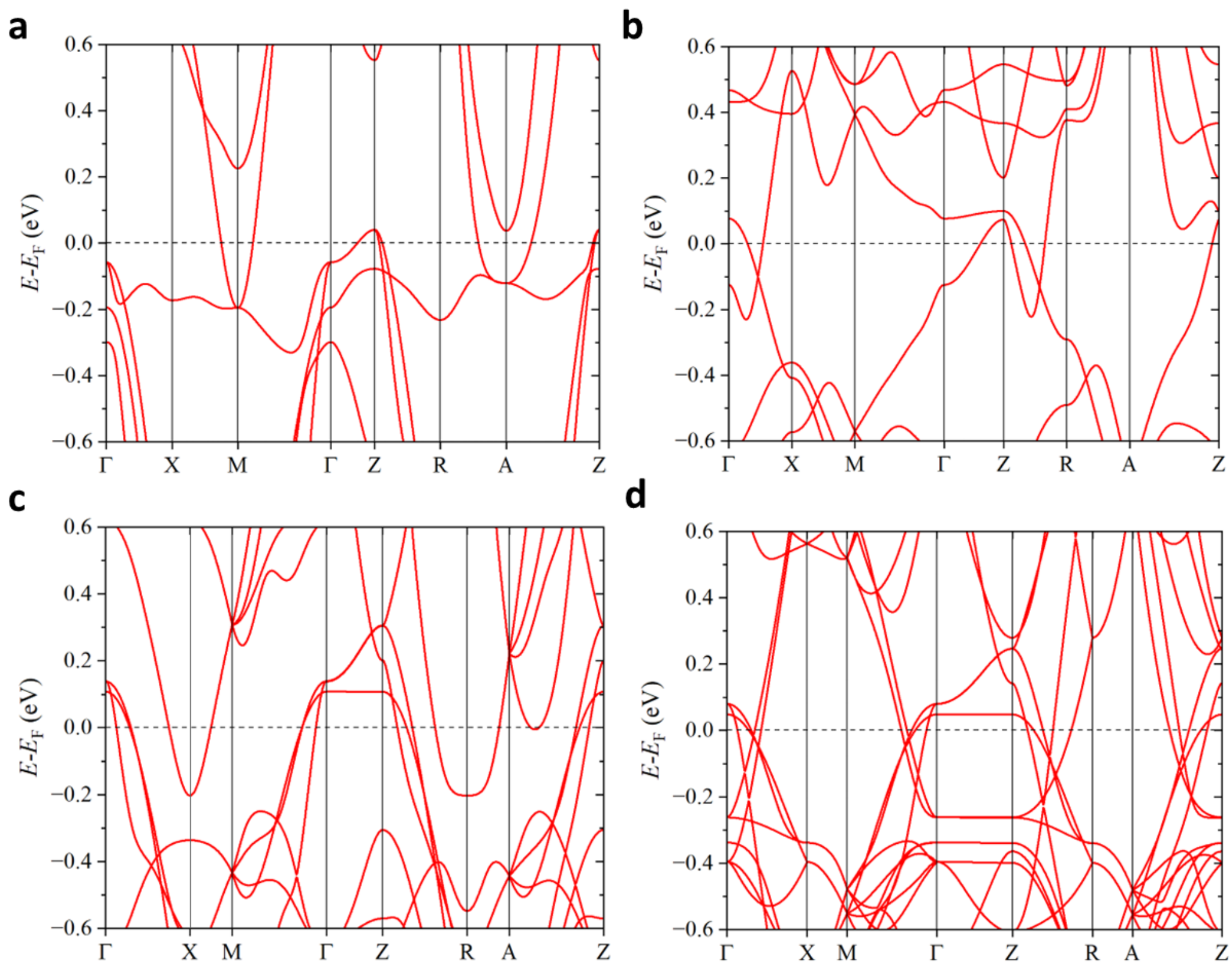

Figure S8. DFT band structures of strained bulk FeTe ($a = b = 3.905$ Å) in different antiferromagnetic configurations: (a) Néel AFM, (b) stripe AFM, (c) bicollinear AFM, and (d) dimer AFM. All band structures are calculated using the corresponding minimal cells of different AFM configurations.

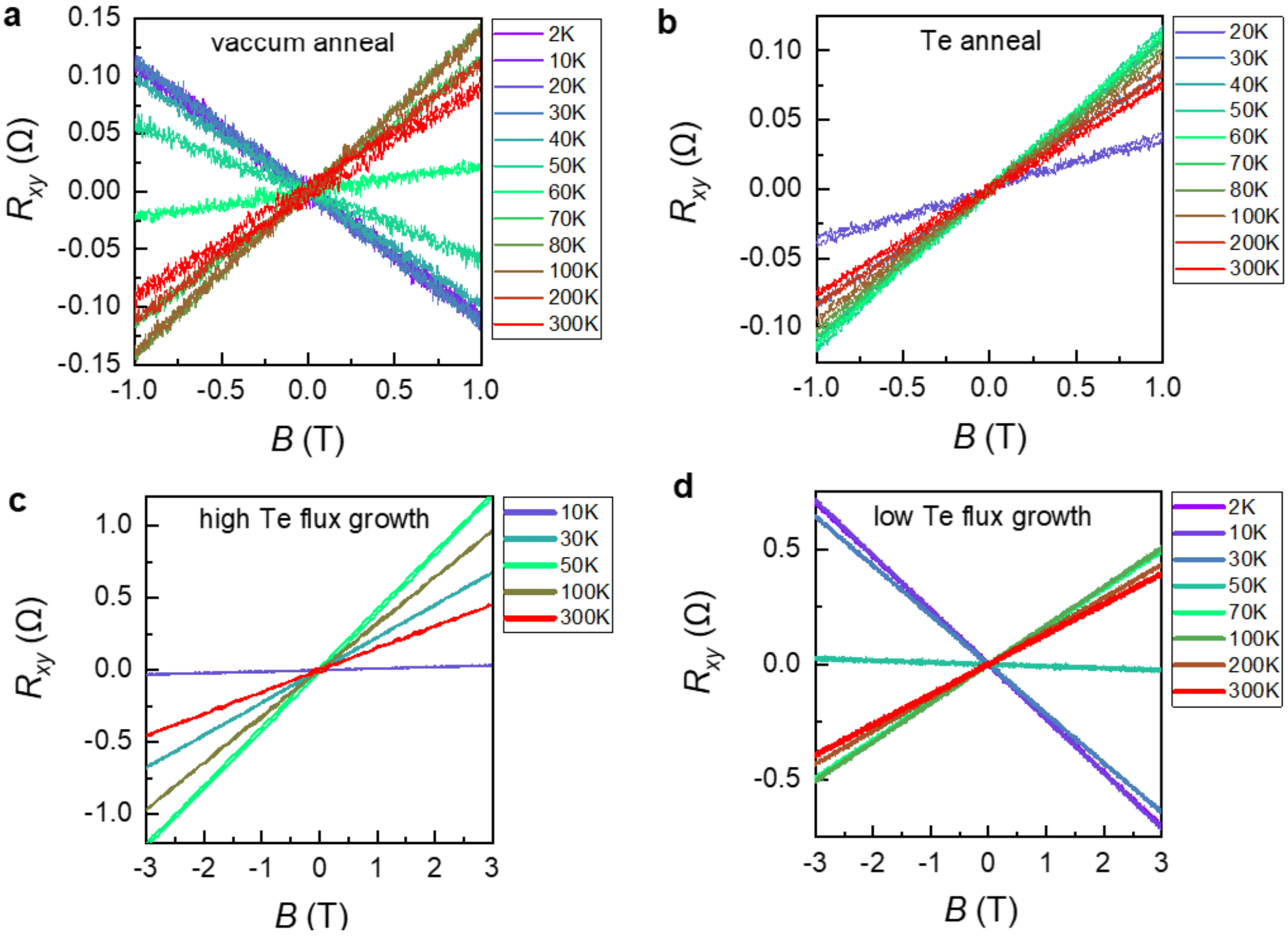

Figure S9. Temperature dependent Hall resistance for the four samples in Figure 5(b).

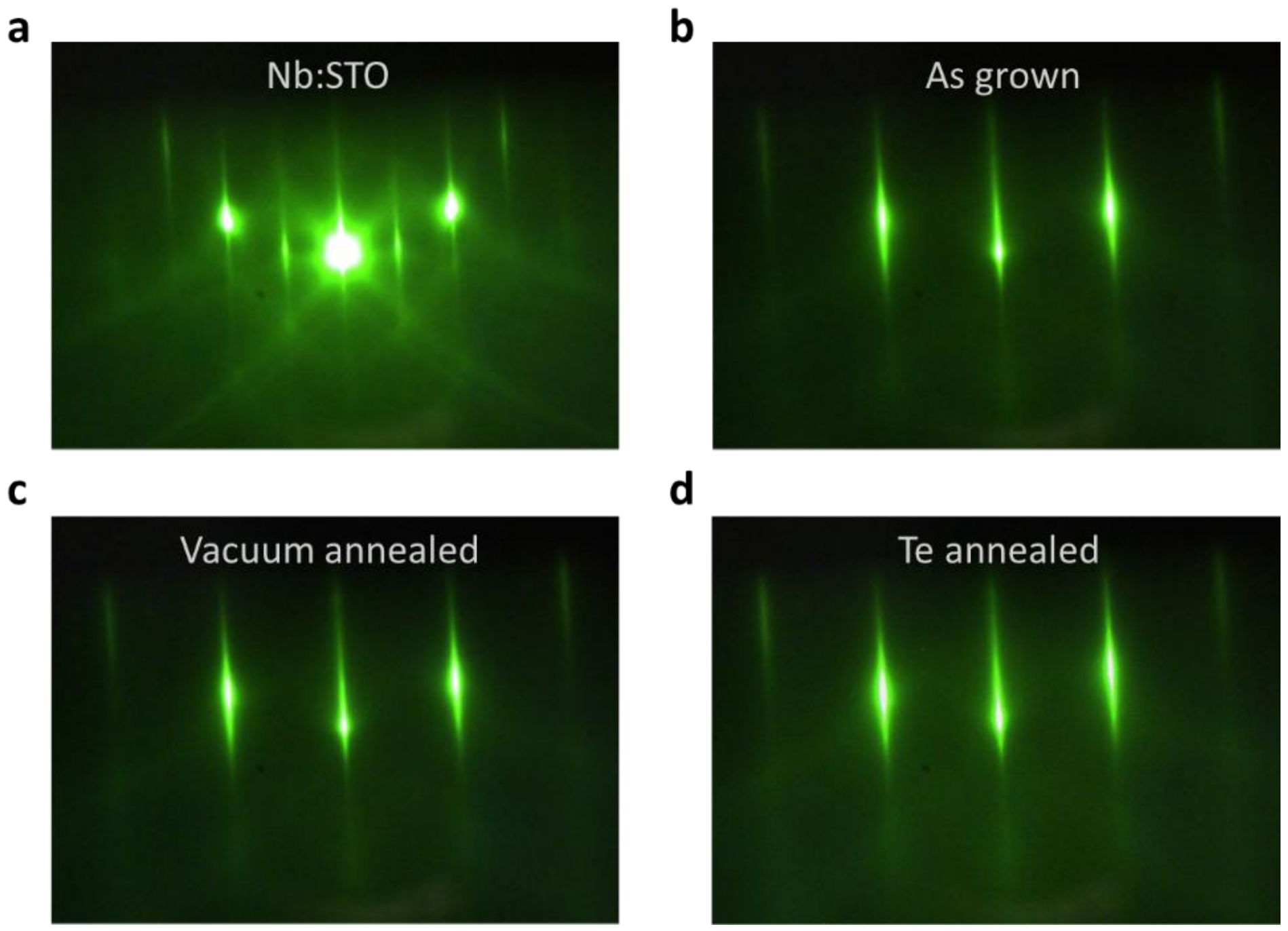

Figure S10. Reflection high-energy electron diffraction (RHEED) images. (a) Nb:STO substrate before growth. (b) As-grown 10-ML FeTe film. (c) The same sample after vacuum annealing for 45 minutes. (d) The same sample after Te-vapor annealing for 15 minutes.